\documentclass[preprint,showpacs,preprintnumbers,amsmath,amssymb]{revtex4-1}
\input{epsf}

\usepackage{graphicx}
\usepackage{dcolumn}

\begin{document}


\title{A proposal for factorization using Kerr nonlinearities between
three harmonic oscillators}
\author{H. T. Ng${}^1$ and Franco Nori${}^{1,2}$}
\affiliation{${}^{1}$Advanced Science Institute, RIKEN, Wako-shi, Saitama 351-0198, Japan\\
${}^2$Physics Department, The University of Michigan, Ann Arbor, Michigan 48109-1040, USA
}
\date{\today}

\begin{abstract}
We propose an alternative method to factorize an integer by using three harmonic oscillators.  These oscillators are coupled together via specific Kerr nonlinear interactions. 
This method can be applied even if two harmonic oscillators are 
prepared in mixed states. As simple examples, we show how to factorize $N=15$ and 35 using this
approach.  
The effect of dissipation of the harmonic oscillators on the performance of this method is studied.  We also study the realization of nonlinear interactions between the coupled oscillators.  However, the probability of finding
the factors of a number is inversely proportional to its input size.
The probability becomes low when this number is large. 
We discuss the limitations of this approach.
\end{abstract}


\maketitle



\section{Introduction}
Current cryptosystems, such as RSA, depend on factorization of a large integer into a product 
of distinct prime numbers.  
In 1994, Peter Shor proposed \cite{Shor} a quantum algorithm to factorize an integer in polynomial 
time.  
This algorithm gives an exponential speed-up on the best known classical algorithm 
\cite{Ekert,Childs}.  
Implementations of Shor's algorithm have been demonstrated in NMR \cite{Vandersypen} 
and photonic systems \cite{Lu,Lanyon} to factorize $N=15$.   
Recently, a new quantum algorithm \cite{Peng} based on the 
adiabatic theorem has been proposed to factorize integers,
and it has been implemented (for $N=21$) in an NMR experiment \cite{Peng}.  
It is an interesting question to ask whether 
an alternative approach exists, to perform factorization.

In this paper, we propose an alternative method to factorize integers 
which uses three coupled harmonic oscillators. Here we show that Kerr-type nonlinear interactions can be used for factorization.   
This method differs from the conventional quantum-circuit based model \cite{Deutsch0,Barcenco}, 
which uses quantum gates for quantum computation.
However, for the method studied here, the probability of finding factors of a number $N$
is inversely proportional to the input size.
Thus, this approach cannot provide any speed-up compared to classical
algorithms for factoring. Still, it is worth studying alternative
methods for factoring, since perhaps this approach may be improved in the
future.

As a particular example of
nonlinearity, we consider the Kerr effect.
Kerr nonlinearities
have been found to be important in quantum optics \cite{Scully}.  
For example, Kerr nonlinearities can be exploited to perform quantum demolition measurements 
\cite{Imoto}, reduce quantum fluctuations of photon numbers \cite{Milburn}, 
and produce Schr\"{o}dinger-cat states \cite{Yurke,Leggett}.
To perform factorization using this approach,  it is necessary to engineer specific Kerr nonlinear
interactions between the three coupled harmonic oscillators.  Recently, the generation of nonlinear interactions 
of a harmonic oscillator has been proposed by coupling it
to a qubit with a time-dependent drive \cite{Jacobs}.  This method can be applied to various
coupled systems \cite{Jacobs}, such as nanomechanical resonators coupled to a 
Cooper-pair box.  In addition, ``three-body''  interactions can be produced in cold atoms \cite{Johnson,Will}
or polar molecules \cite {Buchler} trapped in an optical lattice.  These ``three-body''  interactions can be 
adjusted by external fields \cite{Johnson,Buchler}.   These may pave the way to realize 
Kerr nonlinear interactions between the coupled oscillators.

In our method, an integer $n$ is represented as a number state 
$|n\rangle$  of a harmonic oscillator.   
The states of the first two harmonic oscillators are prepared in a number-state basis,
which encode the trial factors of the number $N$ to be 
factorized.   The third oscillator is prepared in a coherent state.  
We will show that this approach can be applied even 
if the first two harmonic oscillators are in mixed states.  
The quantum coherence of the third harmonic oscillator 
is essential in this algorithm.  

Now we outline the basic idea of this approach.  
We consider a model with specific Kerr nonlinear interactions 
between the three harmonic oscillators.  In the phase-space representation, 
such nonlinear interactions make a coherent state rotate with a ``phase angle'' for a time $t$, 
where this coherent state and the number states of the two trial factors
are in product states.  This ``phase angle'' is proportional to the product of these two trial factors. 
The third oscillator thus acts as a ``marker''  for factors and non-factors.   
By quantum parallelism \cite{Deutsch}, the ``phase angles'' can be simultaneously 
computed for all trial factors.  By performing a conditional measurement of 
a coherent state with a ``phase angle'', which is proportional to the 
product of two factors (i.e., the number $N$ to be factorized),
the resulting state of the oscillators 1 and 2 is the state of the factors. 
The prime and composite factors are obtained after this step.
By using this procedure for factoring the composite factors, the required prime factors of the composite factors can be obtained.

Although, all products of any two trial factors can be simultaneously computed, and they are ``written'', in a single step, to the rotation frequencies of coherent states, the probability of obtaining the required
coherent states is inversely proportional to input size of products of the trial 
factors.  Therefore, the
probability of finding the factors of an integer becomes low when
this integer is large. In addition, it requires a huge amount of energy to encode a number onto a harmonic oscillator when this number is large.

This paper is organized as follows:  Sec.~II, we introduce the system of
three coupled harmonic oscillators.  Sec.~III, we present a method for
factoring integers using these three oscillators.  
Sec.~IV, we show how to factorize a number with an
initial thermal state.  Sec.~V, we investigate the effect of dissipation
to factor a number.  Sec.~VI, we study how 
to realize the specific Kerr nonlinear interactions, with details 
given in Appendix A.  In Sec.~VII, we discuss the limitations of this approach.
Finally, we close this paper
with a conclusion.


\section{System}
We employ three coupled harmonic oscillators to perform quantum factorization.
Its Hamiltonian is
\begin{eqnarray}
\label{Hamiltonian}
H&=&\hbar\sum^3_{j=1}{\omega_j}a^\dag_ja_j+{\hbar}\sum^{K}_{k=1}g_k(a^\dag_1a_1a^\dag_2a_2)^ka^\dag_3a_3,
\end{eqnarray}
where $a_j$ and $\omega_j$ are the annihilation operator and the frequency 
of the $j$-th harmonic oscillator, respectively.  The parameter $g_k$ is 
the coupling strength of the nonlinear 
interaction $(a^\dag_1a_1a^\dag_2a_2)^ka^\dag_3a_3$  which is the product of the number operators 
of all three harmonic oscillators, and the exponent $k$ is a positive integer.

Obviously, this number-conserving Hamiltonian $H$ is exactly solvable. 
The product of number states of the three harmonic oscillators $|n,m,l\rangle$,
\begin{equation}
|n,m,l\rangle=|n\rangle_1|m\rangle_2|l\rangle_3,
\end{equation}
is an eigenstate of $H$ with eigenvalue $E_{n,m,l}$,
\begin{equation}
E_{n,m,l}=\hbar\bigg[{\omega_1n+\omega_2m+\omega_3l+\sum^{K}_{k=1}(nm)^k}l\bigg].
\end{equation}

\section{Factorization}
\begin{figure}[ht]
\centering
\includegraphics[height=3.2cm]{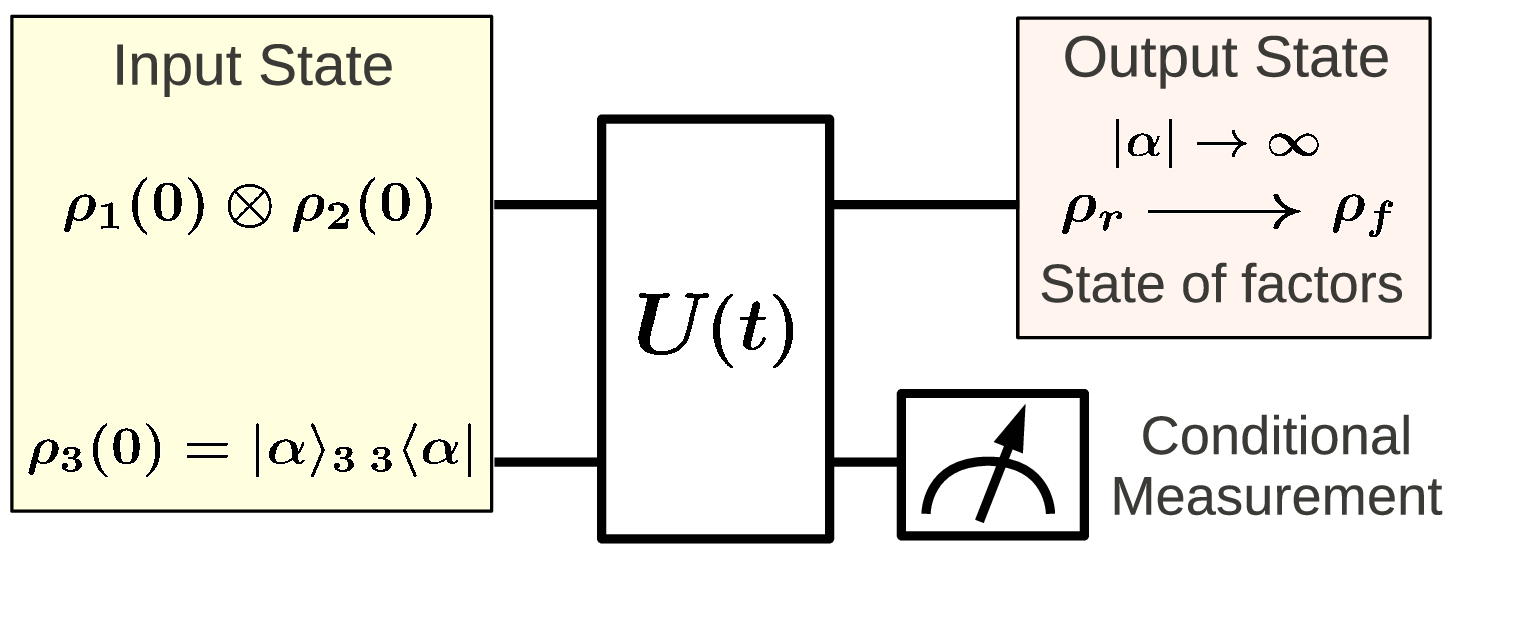}
\caption{ \label{qcircuit} 
(Color online) Factorization using three harmonic oscillators.  We prepare an input state 
$\displaystyle{\rho(0)=\rho_1(0)\otimes\rho_2(0)\otimes\rho_3(0)}$, 
where $\rho_1(0)\otimes\rho_2(0)=\sum_{n,m}{p}^{n'm'}_{nm}|n,m\rangle\langle{n',m'}|$, 
and $\rho_3(0)=|\alpha\rangle_3\,{}_3\langle\alpha|$.
After applying a unitary operator $U(t)$, the reduced density matrix $\rho_r$ can be obtained 
by a conditional measurement of the coherent state $|\alpha_N(t)\rangle_3$, 
which rotates in phase space with a frequency $\Omega_N$.  In the large-$|\alpha|$ limit, 
the reduced density matrix $\rho_r$ becomes the state $\rho_f$ for the factors.}
\end{figure}

Now we present a method to factorize an integer $N$.  
This approach involves only three steps:
initialization, time evolution, and conditional measurement.  
Figure \ref{qcircuit} shows the quantum circuit for this factoring method.

\subsection{Initialization}
The first two harmonic oscillators are prepared in either
pure or mixed states, and 
the third harmonic oscillator is initialized as a coherent state, i.e., 
\begin{eqnarray}
 \rho(0)&=&\rho_{1}(0)\otimes\rho_2(0)\otimes\rho_3(0),\\
\rho_{1}(0)\otimes\rho_2(0)&=&\sum_{n,n',m,m'}p^{n'm'}_{nm}|n,m\rangle\langle{n'},{m'}|,\\
\rho_3(0)&=&|\alpha\rangle_3\,{}_3\langle\alpha|,
\end{eqnarray}
where 
$p^{n'm'}_{nm}$ are the probabilities of the number states $|n,m\rangle\langle{n'},{m'}|$ of 
the oscillators 1 and 2, and $n,n',m,m'=2,\ldots,{\lceil}{N/2}{\rceil}$. 
The trial factors are encoded to the state of the harmonic
oscillators 1 and 2, where each number state represents a trial factor.  
The eigenvalues $nm$ of the state $|n,m\rangle$ are the product of the 
two trial factors $n$ and $m$.

\subsection{Time evolution}
Let us now study the time evolution of 
the system starting with a product state $|n,m\rangle|\alpha\rangle_3$.
It is convenient to work in the interaction picture.  We consider
the unitary transformation:
\begin{eqnarray}
\label{intU}
U'(t)&=&\exp\Big[{-i\Big(\hbar\sum^3_{j=1}{\omega_j}a^\dag_ja_jt\Big)}\Big].
\end{eqnarray}
By applying the unitary operator $U'(t)$ in Eq.~(\ref{intU}) to the Hamiltonian $H$ in Eq.~(\ref{Hamiltonian}), the Hamiltonian is then transformed as
\begin{eqnarray}
H'&=&{\hbar}\sum^{K}_{k=1}g_k(a^\dag_1a_1a^\dag_2a_2)^ka^\dag_3a_3.
\end{eqnarray}
The time-evolution operator $U(t)=\exp{(-iH't)}$ 
transforms the state $|n,m\rangle|\alpha\rangle_3$ into
\begin{equation}
\label{U_inputstate}
U(t)|n,m\rangle|\alpha\rangle_3=|n,m\rangle|\alpha_{nm}(t)\rangle_3,
\end{equation}
where 
\begin{equation}
\alpha_{nm}(t)\!=\!\exp{(-i\Omega_{nm}t)}\alpha, 
\end{equation}
and 
\begin{equation}
\Omega_{nm}\!=\!\sum^{K}_{k=1}g_k(nm)^k
\end{equation}
 is the rotation frequency, in phase space,
of the coherent state $|\alpha_{nm}(t)\rangle$.
Here we have used 
\begin{equation}
\exp(i\vartheta{a}^\dag_3a_3)|\alpha\rangle_3=|\alpha\exp(i\vartheta)\rangle_3,
\end{equation}
where $\vartheta$ is a real number \cite{Barnett}.
The rotation frequency of the coherent state
$|\alpha_{nm}(t)\rangle_3$ is increased by ${\sum^{K}_{k=1}g_k(nm)^k}$, due to 
the nonlinear interaction.
If the product $nm$ is equal to $N$, the coherent state $|\alpha_N(t)\rangle_3$ 
rotates with a frequency 
\begin{equation}
\Omega_N=\sum^{K}_{k=1}g_kN^k,
\end{equation}
in phase space.
Otherwise, the coherent states for $nm\neq{N}$ rotate with frequencies $\Omega_{nm}$,
which are different to the frequency $\Omega_N$.
This means that the coherent states corresponding to factors' states and non-factors' states 
have different rotation frequencies in phase space.
Thus, the state of the harmonic oscillator 3 acts as a ``marker'' for the states of factors
and non-factors.

Now we apply the time-evolution operator $U(t)$ to the initial state $\rho(0)$.
We then write the density matrix $\rho(t)$ 
of the system as
\begin{equation}
\rho(t)=\sum_{n,m,n',m'}{p}^{n'm'}_{nm}|n,m\rangle|\alpha_{nm}(t)\rangle_3\,{}_3\langle
\alpha_{n'm'}(t)|\langle{n',m'}|,
\end{equation}
where $n,m,n'$ and $m'$ denote all trial factors.
By quantum parallelism \cite{Deutsch}, the functions $\alpha_{nm}(t)$ are simultaneously computed for 
different values of $n$ and $m$.  Obviously, this state becomes non-separable \cite{Werner} between
the three harmonic oscillators.  The quantum entanglement \cite{Horodecki} between the harmonic
oscillators is produced due to 
the nonlinear interactions of the three harmonic oscillators.

\subsection{Conditional measurement}
We then perform a measurement conditional on the coherent state $|\alpha_N(t)\rangle_3$,
where this coherent state and the factor's state are in product states at the time $t$.  
In this way, we can abandon the state of non-factors if the orthogonality of coherent
states is approximately valid.  Here we denote the products $rs$ and $r's'$ to be 
equal to $N$ (i.e., $r{\times}s,r'{\times}s'=N$).   After a conditional measurement, 
the reduced density matrix is 
\begin{eqnarray}
\label{reduced_matrix}
\rho_r(t)&=&\frac{1}{A}{\rm Tr}_3\big\{\mathcal{J}\big[|\alpha_N(t)\rangle_3\;{}_3\langle\alpha_N(t)|\big]\rho(t)\big\},
\end{eqnarray}
where $\mathcal{J}\big[|\alpha_N(t)\rangle_3\;{}_3\langle\alpha_N(t)|\big]$ is the measurement operator as \cite{Wiseman}
\begin{eqnarray}
&&\mathcal{J}\big[|\alpha_N(t)\rangle_3\;{}_3\langle\alpha_N(t)|\big]\rho(t)\nonumber\\
&=&|\alpha_N(t)\rangle_3\;{}_3\langle\alpha_N(t)|\rho(t)\big[|\alpha_N(t)\rangle_3\;{}_3\langle\alpha_N(t)|\big]^\dag,
\end{eqnarray}
and $A$ is the probability of obtaining the coherent state $|\alpha_N(t)\rangle_3$,
\begin{eqnarray}
 A&=&{\rm Tr}\big\{\mathcal{J}\big[|\alpha_N(t)\rangle_3\;{}_3\langle\alpha_N(t)|\big]\rho(t)\big\}.
\end{eqnarray}
From Eq.~(\ref{reduced_matrix}), the resulting reduced density matrix $\rho_r$ can be written as  
\begin{eqnarray}
\label{redrho}
\rho_r(t)&=&\frac{1}{A}\sum_{n,m,n',m'}{p}^{n'm'}_{nm}\epsilon_{nm}\epsilon^*_{n'm'}|n,m\rangle\langle{n',m'}|,
\end{eqnarray}
where $A=\sum_{n,m}{p}^{nm}_{nm}|\epsilon_{nm}|^2$
and the coefficient $\epsilon_{nm}$ 
\cite{Barnett},
is the overlap between the two coherent states $|\alpha_N(t)\rangle$ and 
$|\alpha_{nm}(t)\rangle$,
\begin{equation}
\label{overlap}
\epsilon_{nm}=\exp\big\{-|\alpha|^2\big[1-\exp(i\Omega_{N}t-i\Omega_{nm}t)\big]\big\}.
\end{equation}

In the limit of large $|\alpha|$, the orthogonality of coherent states holds 
approximately, such that the overlap
$\epsilon_{nm}$ tends to zero.
Thus, we can obtain the state of the factors 
\begin{equation}
\lim_{|\alpha|\rightarrow{\infty}}\rho_r=\rho_f,
\end{equation}
where $\rho_f$ is the density matrix of the factors
\begin{equation}
 \rho_f=\frac{1}{A_f}\sum_{r,s,r',s'}{p}^{r's'}_{r,s}|r,s\rangle\langle{r',s'}|,
\end{equation}
and $A_f=\sum_{r,s}{p}^{rs}_{rs}$ is a normalization constant.
Indeed, the reduced density matrix $\rho_r$ is well approximated by the state of factors 
$\rho_f$, if $|\alpha|$ is sufficiently large.
Finally, we can determine the factors of the number $N$ by measuring the state 
of the harmonic oscillators 1 and 2.  

We also remark that the overlap between the two coherent states $\epsilon_{nm}$ 
is a function of time $t$.  To achieve the best approximation 
of the state of the two factors, the overlap between the coherent state $|\alpha_{N}(t)\rangle$ 
and the other coherent states in Eq. (\ref{redrho}) can be minimized
by performing the conditional measurement at an appropriate time $t^*$.
This time $t^*$ depends on the initial probability distribution and the non-linear
interaction strength.  

\section{Example: Initial thermal states}
We consider the initial state of the first two harmonic oscillators
in thermal states and the third harmonic oscillator is a coherent state, i.e., 
\begin{equation}
\label{int_thermal}
\rho(0)=\sum_{n,m}{p}_{1n}p_{2m}|n,m\rangle\langle{n,m}|\otimes|\alpha\rangle_3\,{}_{3}\langle\alpha|,
\end{equation}
where $p_{jn}=[1-e^{-\beta\hbar\omega_j}]e^{-\beta{n_j}\hbar\omega_j}$  
is the probability of the thermal states of 
the $j$-th harmonic oscillator \cite{Barnett}, $\beta=1/k_BT$, $k_B$ is the Boltzmann's constant, 
$T$ is the temperature, and $j=1,2$.  The mean excitation number $\bar{n}_j$ is 
$[e^{\hbar\omega_j/k_B{T}}-1]^{-1}$ for the bosonic mode $a_j$ \cite{Barnett}.
Here we require that the probabilities of 
the number states $|n,m\rangle$ are much smaller than that of the state of factors, where $n$ and 
$m$ are larger than $N$.

\begin{figure}[ht]
\centering
\includegraphics[height=6.0cm]{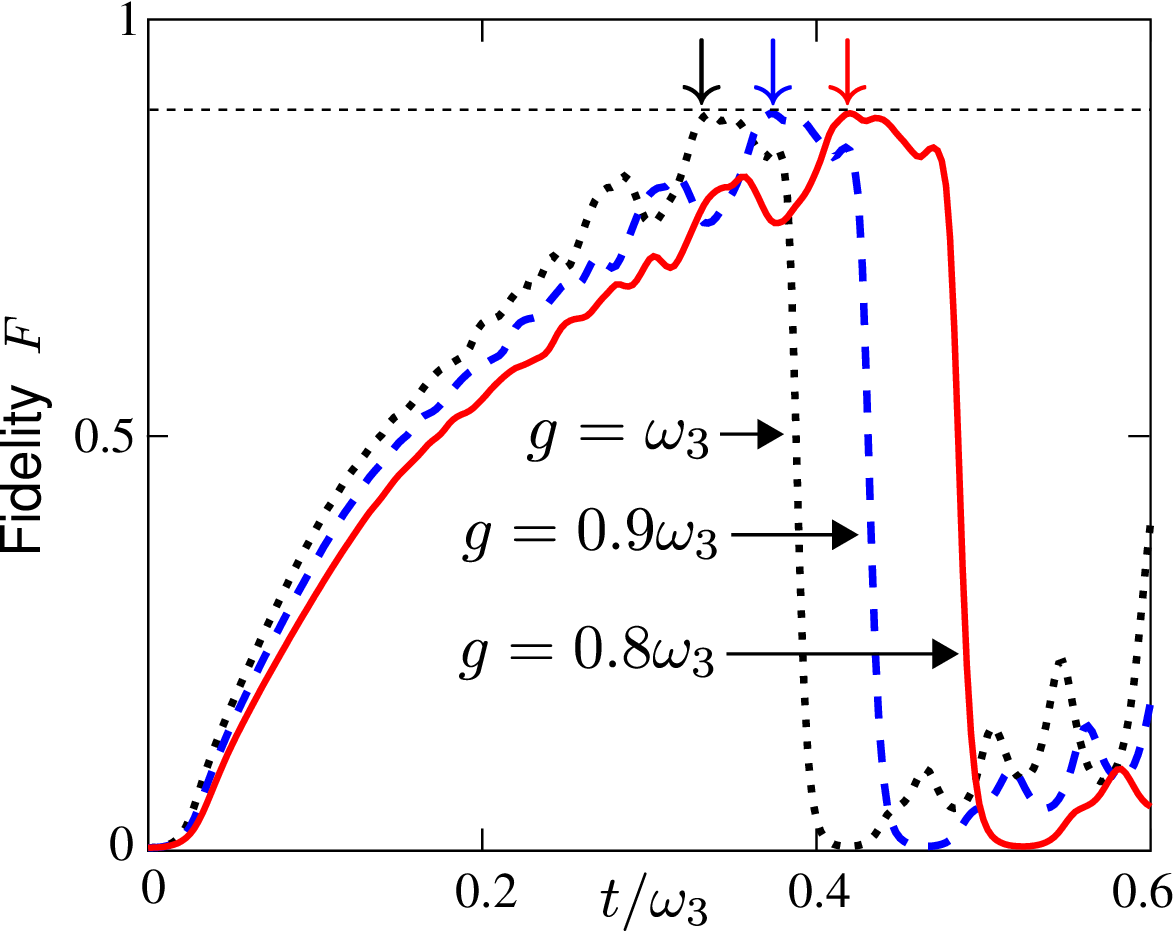}
\caption{ \label{fig_e1}
(Color online) The fidelity $F(t/\omega_3)$ between the reduced density matrix $\rho_r(t/\omega_3)$ and the factor's states $\rho_f$ 
for $N=15$, is plotted as a function of the dimensionless time $t/\omega_3$, 
for the parameters $\omega_1=1.5\omega_3$, $\omega_2=2\omega_3$, $T=3\hbar\omega_3/k_B$ and $|\alpha|=5$.
The mean excitation numbers $\bar{n}_j$ are $1.542$, $1.055$ and $2.528$, for $j=1,2,3$ and $k=1$. 
The probability of obtaining the coherent state $|\alpha_{N=15}(t/\omega_3)\rangle$ is $3.65\times{10}^{-3}$.
Black (dotted), blue (dashed), and red (continuous) lines correspond to the results for $g=\omega_3$, $0.9\omega_3$, and $0.8\omega_3$.
For different strengths $g$, the maxima of the respective fidelities are indicated by the arrows. 
}
\end{figure}

\subsection{Factoring $15$ and $35$}
As a simple illustrative example, we now factor $N=15=3\times{5}$ using this 
quantum approach.  For simplicity, we consider the Hamiltonian 
\begin{equation}
\label{Hamiltoniank}
H_k=\hbar\sum^3_{j=1}{\omega_j}a^\dag_ja_j+{\hbar}g(a^\dag_1a_1a^\dag_2a_2)^ka^\dag_3a_3,
\end{equation}
where $g$ is the nonlinear strength and $k$ is a positive integer.
This allows us to clearly study the role of nonlinearity in this method.

\begin{figure}[ht]
\centering
\includegraphics[height=6.0cm]{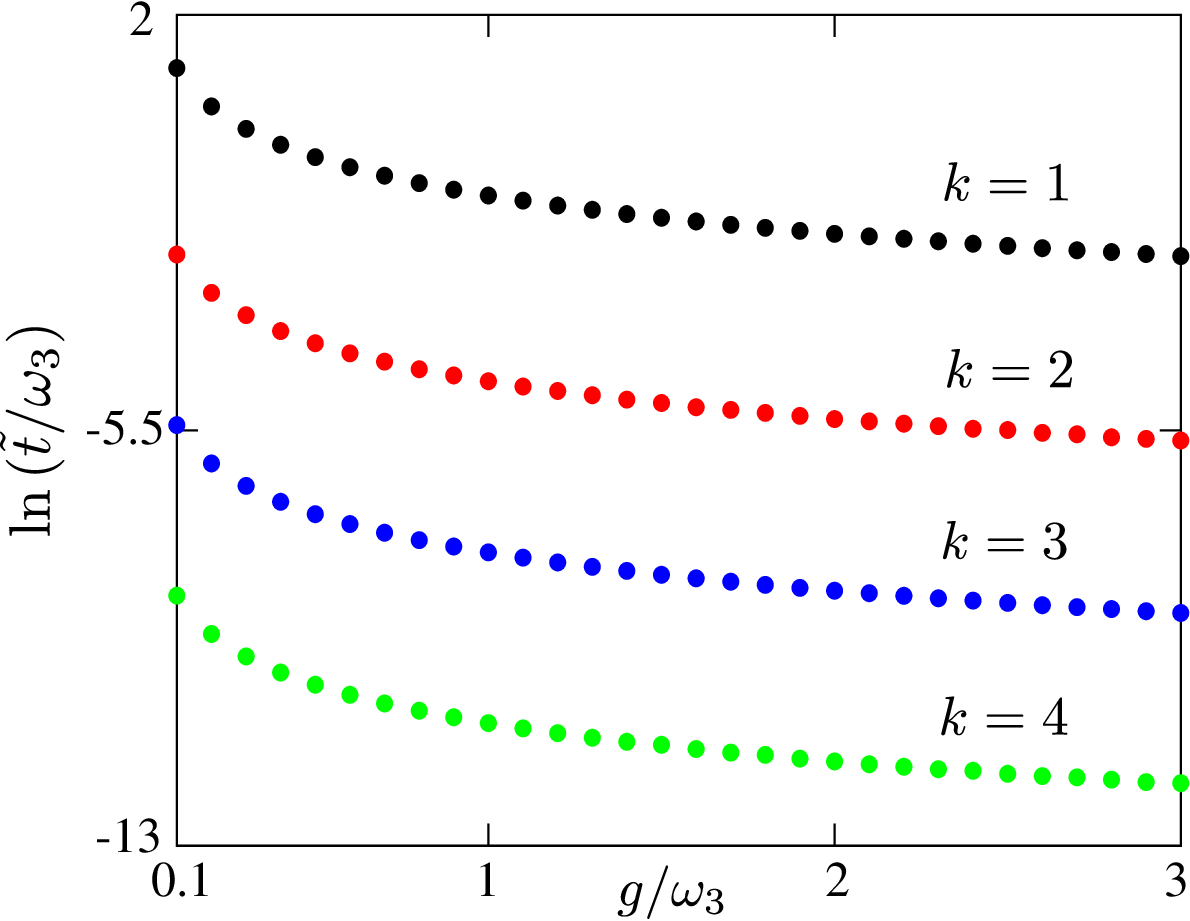}
\caption{ \label{fig_l1}
(Color online) 
The logarithmic dimensionless time $\ln\,(\tilde{t}/\omega_3)$ is plotted as 
a function of the nonlinear strength $g/\omega_3$ for $|\alpha|=6$
and the same parameters as in Fig.~\ref{fig_e1}.
The time $\tilde{t}/\omega_3$ is taken 
when the fidelity $F$ for $N=15$ first arrives at 0.9.
Black (top curve), red, blue, and green (bottom curve), dots correspond to the results for $k=1,2,3$, and 4.
 }
\end{figure}
To evaluate the performance of this approach, 
we investigate the fidelity $F$ between the reduced density matrix $\rho_r(t)$ 
and the state of the factors $\rho_f$ at the time $t$.  
The fidelity is defined as \cite{Uhlmann}
\begin{equation}
 F(t)=\bigg({\rm Tr}\Big\{\Big[{\rho^{1/2}_f\rho_r(t)\rho^{1/2}_f}\Big]\Big\}^{1/2}\bigg)^{2}.
\end{equation}
The fidelity $F(t)$ 
can be obtained for the initial thermal states in Eq. (\ref{int_thermal}), 
\begin{equation}
F(t)={\sum_{r,s}p_{1r}p_{2s}}\bigg({\sum_{n,m}p_{1n}p_{2m}|\epsilon_{nm}|^2}\bigg)^{-1},
\end{equation}
where $r,s=3,5$.
We now take the exponent $k$ in Eq. (\ref{Hamiltoniank}) to be one, i.e., $k=1$.
In Fig. \ref{fig_e1}, the fidelity $F(t/\omega_3)$ of the state of the factors is plotted 
as a function of the dimensionless time $t$ for the magnitude $|\alpha|=5$ 
and different values of the dimensionless nonlinear strengths $g/\omega_3$.  
The fidelity $F$ begins to increase as a function of the time and reaches a 
maximum at $t^*/\omega_3\approx{0.335}$, for $g=\omega_3$.  
Then, the fidelity sharply drops and fluctuates with time $t/\omega_3$.
Fig.~\ref{fig_e1} shows that the fidelities exhibit similar patterns versus $t/\omega_3$ 
for different nonlinear strengths $g$.  Note that, for the smaller strength $g$, 
the fidelity $F$ takes longer time to reach the maximum $(F\simeq{1})$.

We now proceed to study the performance of the algorithm with higher-order $k$
nonlinearity and stronger nonlinear interaction strength $g$.  In Fig.~\ref{fig_l1}, 
we plot the logarithmic dimensionless time $\ln\,(\tilde{t}/\omega_3)$ as a 
function of $g$ for $|\alpha|=6$ and $k=1$.  The 
time $\tilde{t}/\omega_3$ is taken when the fidelity $F$ first arrives at 0.9.  
We can see that the time $\ln\,(\tilde{t}/\omega_3)$ steadily decreases as 
$g/\omega_3$ increases.  In the same figure, we also show results 
for the higher values of $k$~($=2,3$ and 4).  
The logarithmic time decreases by 3 as $k$ increases by 1.  
Thus, a higher-order nonlinearity 
can exponentially shorten the evolution time to achieve the high fidelity $F^*\gtrsim{0.9}$.  
This result shows that both the {\it high-order $k$ nonlinearity and strong nonlinear strength $g$
are very useful to decrease the required time evolution of the system for finding factors}.

\begin{figure}[ht]
\centering
\includegraphics[height=6.0cm]{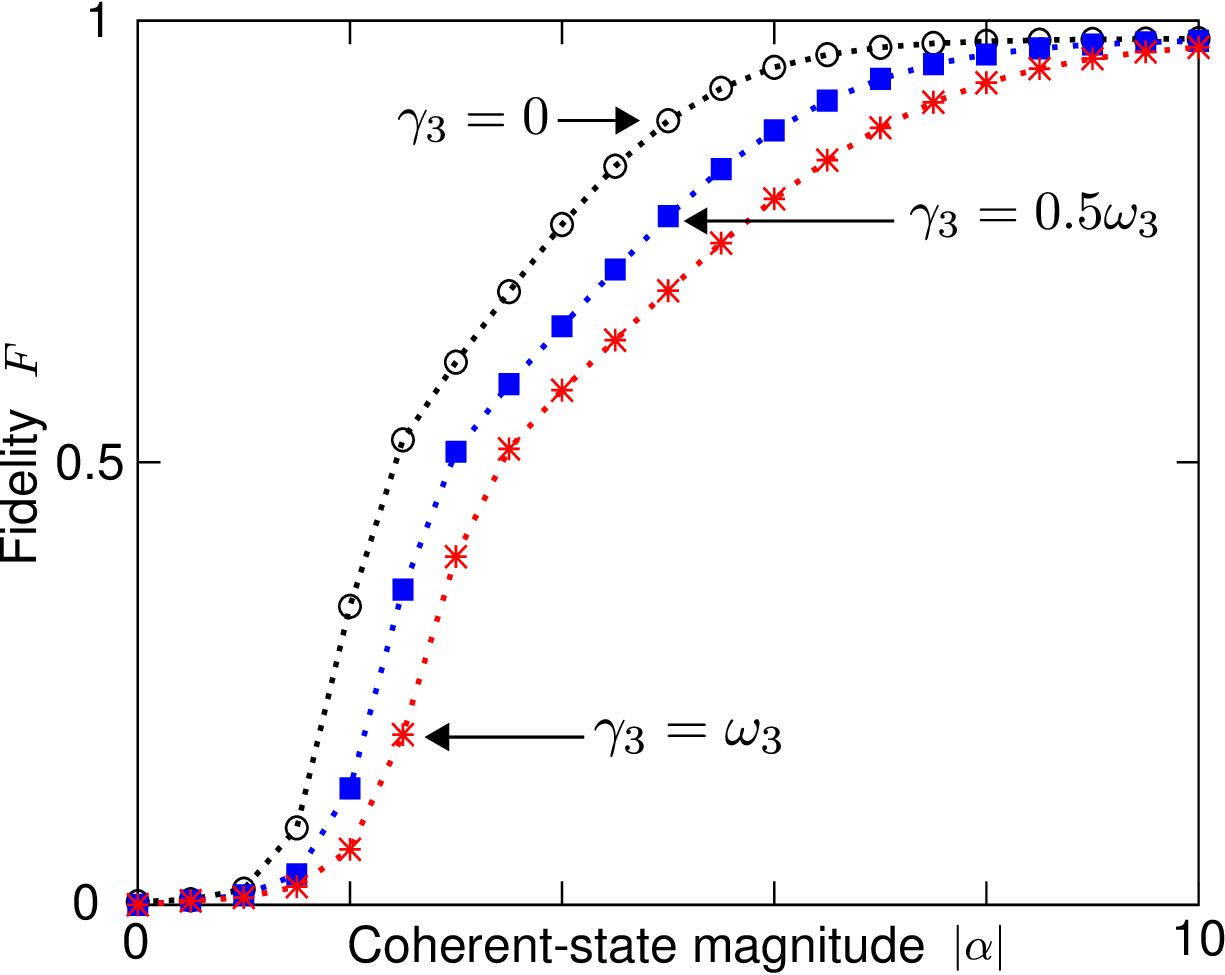}
\caption{ \label{fig_e2}
(Color online) The fidelity $F(t^*/\omega_3)$ for $N=15$, 
plotted as a function of $|\alpha|$ at the time $t^*/\omega_3$, 
for the nonlinear strength $g=\omega_3$ and the same parameters as in Fig.~\ref{fig_e1}.
The fidelities vary with the dissipation rate $\gamma_3$ as shown.  Of course, the fidelity is
higher for zero dissipation, $\gamma_3=0$.}
\end{figure}
In Fig.~\ref{fig_e2}, we plot the fidelity $F(t^*/\omega_3)$ versus 
$|\alpha|$, for $t^*/\omega_3=0.335$ and $g=\omega_3$.
The fidelity $F$ increases as the value of $|\alpha|$
increases.  The fidelity $F$ becomes saturated and reaches near unity when $|\alpha|\approx{7}$.  
Note that the fidelity cannot be exactly equal to one because the probabilities of the 
initial number states $|n,m\rangle$
are not exactly zero 
where $n$ and $m$ are larger than $15$.

\begin{figure}[ht]
\centering
\includegraphics[height=6.0cm]{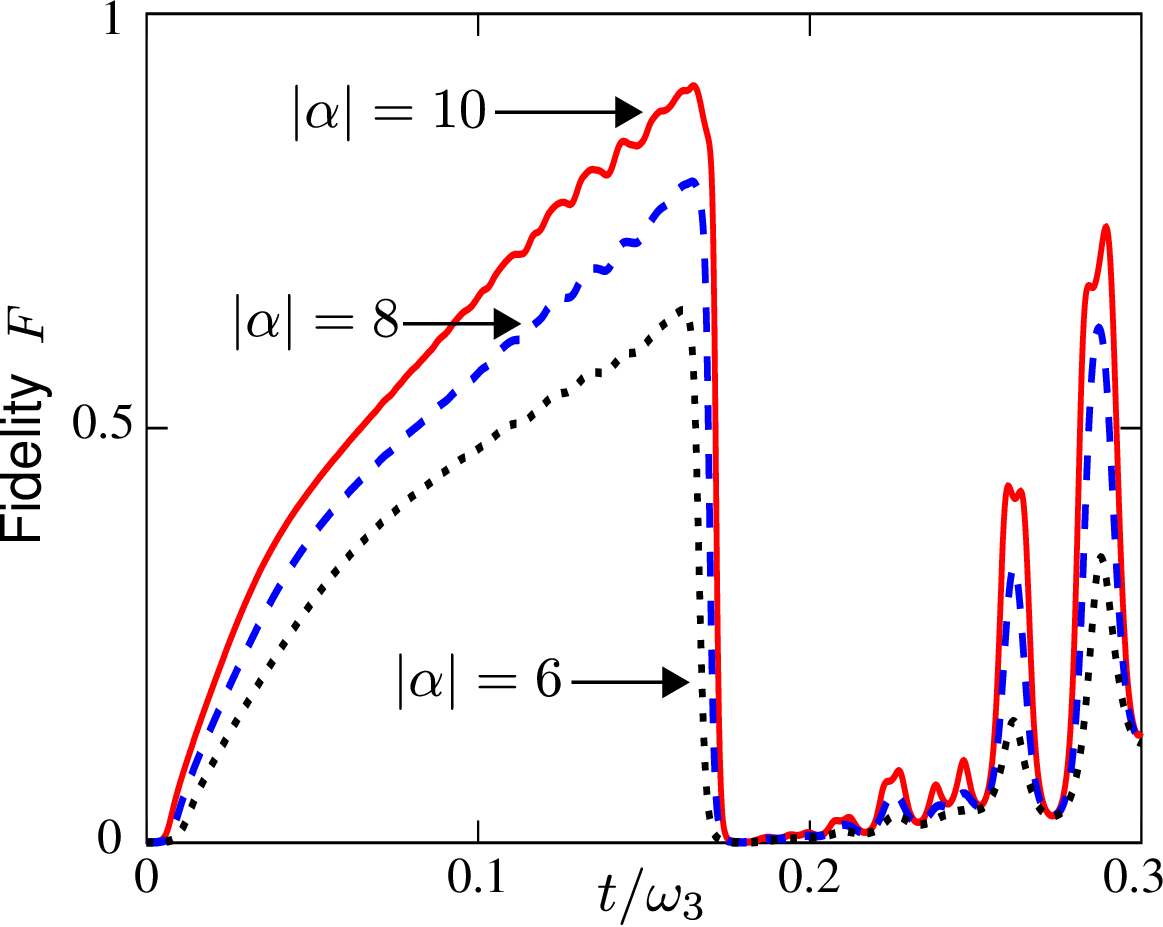}
\caption{ \label{fig_N_35}
(Color online) The fidelity $F(t/\omega_3)$ for $N=35$, 
plotted versus the time $t/\omega_3$, for $g=\omega_3$ and 
the same parameters as in Fig.~\ref{fig_e1}. The probability of obtaining the 
coherent state $|\alpha_{N=35}(t/\omega_3)\rangle$ is $3.54\times{10}^{-4}$.
Black (dotted), blue (dashed), and red (continuous) lines correspond to the results for $|\alpha|=6$, $8$, and $10$.
}
\end{figure}

In addition, this approach can be used for factoring larger integers.
We show another example for factoring $N=35=5\times{7}$.  In Fig.~\ref{fig_N_35},
we plot the fidelity $F(t/\omega_3)$ of the factor's states for $N=35$ versus the time 
$t/\omega_3$.  
High fidelities can be obtained by increasing the magnitude $|\alpha|$ but the 
probability of obtaining the coherent state $|\alpha_{N=35}(t/\omega_3)\rangle$ becomes very low.  
This probability can indeed be increased by appropriately choosing an initial state.

\section{Effect of dissipation}
We now consider the system weakly coupled to a thermal environment.  We also assume that the Markovian
approximation is valid \cite{Breuer}.  The master equation of the density matrix then reads
\begin{equation}
 \dot{\rho}=-i[H',\rho]+\sum^3_{j=1}\big[\mathcal{L}_j(\rho)+\tilde{\mathcal{L}}_j(\rho)\big],
\end{equation}
where 
\begin{eqnarray}
\mathcal{L}_j(\rho)&=&\gamma_j(\bar{n}_j+1)(2a_j\rho{a}^\dag_j-a^\dag_j{a}_j\rho-{\rho}a^\dag_ja_j)/2,\\
\tilde{\mathcal{L}}_j(\rho)&=&\gamma_j\bar{n}_j(2a^\dag_j\rho{a}_j-a_ja^\dag_j\rho-\rho{a_ja^\dag_j})/2,
\end{eqnarray}
and $\gamma_j$ is the damping rate of the $j$-th harmonic oscillator.

Now we consider the dissipative dynamics of the system for the initial state in Eq. (\ref{int_thermal}).
Since the states of the harmonic oscillators 1 and 2 are prepared in thermal states which are
the steady states of harmonic oscillators in the thermal environment,  we can write the total
density matrix $\rho(t)$ at time $t$ as 
\begin{equation}
\rho(t)=\sum_{n,m}p_{1n}p_{2m}|n,m\rangle\langle{n,m}|\otimes\rho^{nm}_3(t),
\end{equation}
where $\rho^{nm}_3(t)$ is the density matrix of the harmonic oscillator 3 corresponding to the number states
$|n,m\rangle\langle{n,m}|$.  Only the harmonic oscillator 3 is subject to 
dissipation.  This allows us to write the master equation of the density matrix $\rho^{nm}_3$ as
\begin{equation}
\label{master3}
 \dot{\rho}^{nm}_3=-i\Omega_{nm}[a^\dag_3{a}_3,\rho^{nm}_3]+\mathcal{L}_j(\rho^{nm}_3)+\tilde{\mathcal{L}}_j(\rho^{nm}_3).
\end{equation}
This master equation is exactly solvable \cite{Daniel}.  
Therefore, the time evolution of the total density matrix $\rho(t)$ 
can be solved by summing over all possible solutions $\rho^{nm}_3(t)$
in Eq.~(\ref{master3}) corresponding to the probabilities $p_{1n}p_{2m}$.

The fidelity can be expressed in terms of the ${Q}$-function as
\begin{eqnarray}
 F(t)&=&\sum_{r,s}p_{1r}p_{2s}Q_{rs}(t)\bigg[\sum_{n,m}p_{1n}p_{2m}Q_{nm}(t)\bigg]^{-1},
\end{eqnarray}
where the ${Q}$-function, $Q_{nm}(t)$, is  \cite{Daniel}
\begin{widetext}
\begin{eqnarray}
&&Q_{nm}(t)\nonumber\\
&=&{}_3\langle{\alpha_N}(t)|\rho^{nm}_3|\alpha_N(t)\rangle_3,\\
&=&\sum^{\infty}_{u,v=0}\frac{[\alpha\alpha^*_N(t)]^u}{u!}\frac{[\alpha_N(t)\alpha^*]^v}{v!}\Lambda^{(u+v+1)/2}_3
\exp[-i\Omega_{nm}(u-v)t+\lambda_0]\exp[(\Lambda_+-1)(\Lambda_--1)|\alpha|^2],\nonumber\\
\end{eqnarray}
\end{widetext}
and 
\begin{eqnarray}
 \Lambda_{\pm}&=&\frac{2\lambda_{\pm}\sinh\phi}{2\phi\cosh\phi-\lambda_3\sinh\phi},\\
\Lambda_{3}&=&\bigg(\cosh\phi-\frac{\lambda_3}{2\phi}\sinh\phi\bigg)^{-2},\\
\phi&=&\bigg(\frac{\lambda^2_3}{4}-\lambda_+\lambda_-\bigg)^{1/2}.
\end{eqnarray}
The parameters $\lambda_{\pm}$, $\lambda_{3}$ and $\lambda_0$ are  
\begin{eqnarray}
\lambda_+&=&\gamma\bar{n}t,~~~~~~~~~~~~\,\lambda_-=\gamma(\bar{n}+1)t,\\
\lambda_3&=&-\lambda_+-\lambda_-,~~~~\lambda_0=\frac{1}{2}(\lambda_--\lambda_+).
\end{eqnarray}

In Fig.~\ref{fig_e2}, we plot the fidelity of the states for the factors versus
$|\alpha|$ subject to the dissipations $\gamma_3=0$ (black circle dotted line), 
$\gamma_3=0.5\omega_3$ (blue square-dotted line) and $\gamma_3=\omega_3$ (red star-dotted line). 
As shown in Fig.~\ref{fig_e2}, the fidelity $F(t^*/\omega_3)$ decreases for small $|\alpha|$
if the dissipation rate $\gamma_3$ is comparable to the nonlinear interaction strength $g$.
However, we can see that the effect of dissipation on the performance of this approach 
is negligible when $|\alpha|$ is about 8.  The high fidelity can be obtained
in the presence of dissipation if the value of $|\alpha|$ is sufficiently large.

\section{Possible implementation for realizing Kerr-type nonlinear interactions}
To perform quantum factorization using this approach, it is
necessary to engineer Kerr-type nonlinear interactions in Eq.~(\ref{Hamiltonian}) 
between the three coupled oscillators.  
Here we briefly discuss several possible ways to realize these nonlinear interactions in Eq.~(\ref{Hamiltonian}),
for $k=1$.  

\subsection{Time-dependent control method: 
coupling harmonic oscillators to a qubit}
Engineering nonlinear interactions of a nanomechanical resonator has been proposed 
by applying a time-dependent drive to a single superconducting qubit 
which is coupled to the resonator \cite{Jacobs}.  Now we generalize this method to produce nonlinear 
interactions between three coupled harmonic oscillators by using a single qubit. 
This method can be applied to different systems such as a superconducting qubit 
coupled to three nanomechanical resonators \cite{Jacobs}, or three different cavity modes 
inside a superconducting resonator \cite{Leek}.  Furthermore, this can be realized by 
a two-level atom coupled to a multi-component atomic 
Bose-Einstein condensate \cite{Ng1}.  

We consider a single qubit dispersively interacting with the three harmonic 
oscillators.  The Hamiltonian $H'$ of this system can be written as
\begin{eqnarray}
\label{Ham_qubho}
 H'&=&H_0'+H_I',\\
 H_0'&=&\hbar\omega_z\sigma_z+\hbar\sum^3_{j=1}{\omega_j}a^\dag_ja_j,\\
 H_I'&=&\hbar\sigma_z\sum^3_{j=1}\chi_ja^\dag_j{a}_j,
\end{eqnarray}
where $\omega_z$ is the energy frequency of the qubit, 
$\chi_j$ is the strength of the nonlinear interaction between the $j$-th harmonic oscillator 
and a qubit, for $j=1,2$ and 3.

According to the Zassenhaus formula \cite{Witschel}, the specific nonlinear interactions can be 
generated by applying an appropriate sequence of rotations to the qubit.
It is convenient to work in the interaction picture.  
Here the exponential operators $e^{X_j+Y_j}$ can be written as, 
based on the Zassenhaus formula \cite{Witschel},  
\begin{equation}
\label{Zassenhaus1}
 e^{X_j+Y_j}=e^{X_j}e^{Y_j}e^{Z^{(1)}_j}e^{Z^{(2)}_j}\dots,
\end{equation}
where $Z^{(1)}_j$ and $Z^{(2)}_j$ are the second- and the third-order terms as \cite{Witschel}
\begin{eqnarray}
\label{Z2nd}
Z^{(1)}_j&=&-\frac{1}{2}[X_j,Y_j],\\
\label{Z3rd}
Z^{(2)}_j&=&\frac{1}{3}[Y_j,[X_j,Y_j]]+\frac{1}{6}[X_j[X_j,Y_j]].
\end{eqnarray}
By putting $X_j=-i\sigma_yB_j\tau$ and $Y_j=-i\sigma_zB_j\tau$,
the three-mode Kerr nonlinear interactions can be created up to the third-order expansion
in Eqs.~(\ref{Zassenhaus1}) and (\ref{Z3rd}), where the operators 
$B_j$ from Eqs.~(\ref{B1}) to (\ref{B7}) are sums of the number operators 
$a^\dag_i{a}_i$, for $i=1,2$ and 3.

The derivation of the resulting evolution operator is given in Appendix A.
The resulting evolution operator $\exp({Z})$ can be obtained, where $Z$ is
\begin{eqnarray}
\label{Z}
 Z&=&i\hbar{\tilde{\chi}}a^\dag_1{a}_1a^\dag_2a_2a^\dag_3a_3,
\end{eqnarray}
and 
\begin{equation}
\tilde{\chi}{\approx}12.65\chi_1\chi_2\chi_3\tau^3.  
\end{equation}
Here we have omitted a constant in Eq.~(\ref{Z}) [see Eq.~(\ref{appZ}) in Appendix A].
In this way, the nonlinear 
interactions between the coupled oscillators can be  produced.   
Of course, higher accuracy can be achieved if the time interval $\tau$ is shorter.  
For example, the time interval $\tau$ can be made around a nanosecond for
a Cooper-pair box coupled to a nanomechanical resonator \cite{Jacobs}.  
Note that the coupling strengths $\chi_j$ are much larger
than the damping rate of the resonators \cite{Jacobs}.  
By repeatedly applying the sequence in Eq.~(\ref{int_sequence}),
the total time duration $t$ is sufficiently long to obtain the
high fidelity of the states of the factors.

The harmonic oscillators can also be coupled to a more complex system.  
For example, two degenerate cavity modes with two different polarizations coupled
to an ensemble of two-level atoms can be used for generating cross-Kerr nonlinearities \cite{Brandao}.  
Moreover, the cross-Kerr nonlinear interactions between two modes \cite{Kumar}, 
which have slightly different frequencies, can be generated in a nonlinear-superconducting ring resonator.  
Those systems can be generalized to realize nonlinear interactions between
oscillators.  However, the detailed studies of these proposals will not be presented in
this paper.

\subsection{Atoms or polar molecules in optical lattices}
Effective ``three-body'' interactions \cite{Johnson} can be produced due to the two-body atomic collisions
in optical lattices.  An effective Hamiltonian of bosonic atoms in a single site, 
up to the second-order perturbation, can be written as \cite{Johnson}
\begin{equation}
 H_{\rm eff}=\frac{U_2}{2}n(n-1)+\frac{U_3}{6}n(n-1)(n-2),
\end{equation}
where $U_2$ and $U_3$ are the two-body and three-body interaction energies,
and $n$ is the number operator of atoms in the ground vibrational mode. 
These ``three-body'' interactions have been observed in the revivals of
quantum phase of atoms in a single lattice site \cite{Will}.

Three-mode Kerr nonlinear interactions between the oscillators 
can be realized using 
multi-species atomic condensates trapped in an optical lattice.
The strength of interactions between the atoms can be adjusted 
by Feshbach resonance or by changing the depth of the potential well \cite{Johnson}.  
However, these effective ``three-body'' interactions between cold atoms 
are contributed from the second-order processes of the two-body interactions.  
The three-body interactions are relatively weak compared 
to two-body interactions.  
 
Another promising candidate is polar molecules in optical lattices \cite{Buchler}. 
Polar molecules have large dipole moments and each molecule has a complex internal structure. 
The two-body and three-body interactions of polar molecules can be 
independently tuned \cite{Buchler}.  In this way, controllable nonlinear interactions can be produced between the
molecules.  However, the number of either cold atoms or polar molecules trapped in each lattice site is 
relatively small.  A few tens to several hundreds of atoms or molecules can be trapped 
in each lattice site.

\section{Limitations of this approach}
In this section, we discuss the limitations of this approach.
Let us summarize the two main limitations:
\begin{enumerate}
  \item The probability of obtaining the coherent
state $|\alpha_N(t)\rangle$ is inversely proportional to the input
size of the products of trial factors.
This probability becomes low when the number $N$ is large.
For example, this probability is proportional to $N^{-3/2}$ if the initial
uniform distribution of a pure state is used.  This initial pure state of the first two harmonic oscillators
is prepared as the superposition of number states, 
\begin{equation}
 |\Psi(0)\rangle=\frac{1}{C}\sum^{{\lceil}\sqrt{N}\!~{\rceil}}_{n=3}\sum^{{\lceil}{N/3}{\rceil}}_{m=\lceil\sqrt{N+1}\!~\rceil}|n\rangle_1|m\rangle_2,
\end{equation}
where 
\begin{equation}
 C=({\big{\lceil}\sqrt{N}\!~\big{\rceil}-2})^{1/2}({\big{\lceil}{N/3}\big{\rceil}-\lceil\sqrt{N+1}\!~\rceil}+1)^{1/2}
\end{equation}
is a normalization constant, while $n$ is a number from 3 to ${\lceil}{\sqrt{N}}{\rceil}$, and $m$ is a number from ${\lceil}{N/3}{\rceil}$
and $\lceil\sqrt{N+1}\!~\rceil$.  The symbol ${\lceil}{\sqrt{N}}{\rceil}$
denotes the closest integer $\sqrt{N}$ and larger than $\sqrt{N}$.
  \item It requires an exponential amount of energy to encode
a number $N$ onto a harmonic oscillator.  For example, to encode
a number $N$, an energy $\hbar\omega{N}$ is required, where $\omega$
is the frequency of a harmonic oscillator. The energy resource
becomes large when the encoded number is large.  The required 
energy may be reduced if a better encoding method is used.    
\end{enumerate}

\section{conclusion}
In summary, we have proposed an alternative approach to factor 
integers using the Kerr nonlinearities of three coupled harmonic oscillators.  
This method can work if the harmonic oscillators 1 and 2 are in mixed states.  
The time evolution for finding factors significantly decreases if a higher-order nonlinearity
and stronger nonlinear strength are used.  The effect of
dissipation on the performance of this factoring approach was also studied. 
We have discussed how to realize the specific Kerr nonlinear 
interactions between the three coupled harmonic oscillators.
But the probability of finding the factors of a number $N$ 
becomes low when the number $N$ is large.
The limitations of this approach have also been discussed.




\begin{acknowledgments}
We thank D. Browne and A. Miranowicz for useful comments.
FN acknowledges partial support from the Army Research Office, 
JSPS-RFBR contract No.~09-02-92114, 
Grant-in-Aid for Scientific Research (S), 
MEXT Kakenhi on Quantum Cybernetics, and the
Funding Program for Innovative R\&D on Science and Technology  (FIRST).
\end{acknowledgments}

\newpage
\appendix
\section{Derivation of the evolution operator for nonlinear interactions}
We now present the details of deriving the resulting evolution operator for
nonlinear interactions in Eq.~(\ref{Hamiltonian}), for $k=1$. 
The nonlinear interations can be produced by applying a sequence of rapid rotations to
the qubit, where the qubit dispersively interacts with the three harmonic 
oscillators as described by the Hamiltonian $H'$ in Eq.~(\ref{Ham_qubho}).

To generate the nonlinear interactions in Eq.~(\ref{Hamiltonian}) for
$k=1$, we need to cancel the first- and second-order terms in the Zassenhaus formula.
Based on this formula, the exponential
operators $e^{-Y_j+X_j}$ can be written as \cite{Witschel}  
\begin{equation}
 e^{-Y_j+X_j}=e^{-Y_j}e^{X_j}e^{\tilde{Z}^{(1)}_j}e^{\tilde{Z}^{(2)}_j}\dots,
\end{equation}
where $\tilde{Z}^{(1)}_j$ and $\tilde{Z}^{(2)}_j$ are the second- and the third-order terms as
\begin{eqnarray}
\tilde{Z}^{(1)}_j&=&-\frac{1}{2}[-Y_j,X_j]=Z^{(1)}_1,\\
\tilde{Z}^{(2)}_j&=&\frac{1}{3}[X_j,[X_j,Y_j]]-\frac{1}{6}[Y_j,[X_j,Y_j]].
\end{eqnarray}
By considering the inverse of the exponential operator $e^{-Y_j+X_j}$, 
the exponential operator $e^{-(X_j-Y_j)}$ is
\begin{eqnarray}
\label{Zassenhaus2}
e^{-(X_j-Y_j)}=e^{-\tilde{Z}^{(2)}_j}e^{-\tilde{Z}^{(1)}_j}e^{-X_j}e^{Y_j}\dots.
\end{eqnarray}

From Eqs.~(\ref{Zassenhaus1}) and (\ref{Zassenhaus2}), 
the third-order terms can be obtained by considering the following sequence:
\begin{eqnarray}
\label{3ordersequence}
e^{-(X_j-Y_j)}e^{-Y_j}e^{X_j}e^{-Y_j}e^{-X_j}e^{X_j+Y_j}&=&e^{Z^{(2)}_j-\tilde{Z}^{(2)}_j}.~~~~
\end{eqnarray}

The qubit can be rotated with an arbitrary angle by applying the qubit drive as \cite{Jacobs}
\begin{eqnarray}
\label{qdrive1}
e^{i\theta_j\sigma_x/2}{\sigma_y}e^{-i\theta_j\sigma_x/2}=\sigma_y\cos\theta_j-\sigma_z\sin\theta_j,\\
\label{qdrive2}
e^{i\theta_j\sigma_x/2}{\sigma_z}e^{-i\theta_j\sigma_x/2}=\sigma_z\cos\theta_j+\sigma_y\sin\theta_j,
\end{eqnarray}
By applying the sequence of rotations to the system 
as: 
$\theta_1=\pi/4$, $\theta_2=-\pi/2$, $\theta_3=-\pi$,
$\theta_4=\pi/2$, $\theta_5=-\pi$ and $\theta_6=-\pi/4$,
we can obtain $X_j$ and $Y_j$ in Eq.~(\ref{3ordersequence}),
\begin{eqnarray}
 X_j&=&-i\sigma_yB_j\tau,~~~~Y_j=-i\sigma_zB_j\tau,
\end{eqnarray}
where $\theta_j$ is the angle of the $j$-th rotation for $j=1,\ldots,6$.
The third-order terms can be obtained as
\begin{eqnarray}
Z^{(2)}_j-\tilde{Z}^{(2)}_j=-2iB^3_j\tau^3\bigg[\sigma_y'+\frac{\sigma_z'}{3}\bigg].
\end{eqnarray}

We have assumed that the coupling between the qubit and the individual oscillator
can be switched on and off.
The required nonlinear interactions can be achieved by applying the following sequence:
\begin{eqnarray}
\label{int_sequence}
&&\prod^{7}_{j=1}e^{-(X_j-Y_j)}e^{-Y_j}e^{X_j}e^{-Y_j}e^{-X_j}e^{X_j+Y_j}\nonumber\\
&=&\prod^{7}_{j=1}e^{Z^{(2)}_j-\tilde{Z}^{(2)}_j},
\end{eqnarray}
where
\begin{eqnarray}
\label{B1}
B_1&=&(\chi_1{n}_1+\chi_2n_2+\chi_3n_3)\tau,\\
B_2&=&-(\chi_1{n}_1+\chi_2n_2)\tau,\\
B_3&=&-(\chi_1n_1+\chi_3n_3)\tau,\\
B_4&=&-(\chi_2n_2+\chi_3n_3)\tau,\\
B_5&=&\chi_1n_1\tau,\\
B_6&=&\chi_2n_2\tau,\\
\label{B7}
B_7&=&\chi_3n_3\tau,
\end{eqnarray}
and $n_i=a^\dag_i{a}_i$, for $i=1,\ldots,3$.
Then, the qubit drive in Eqs.~(\ref{qdrive1}) and (\ref{qdrive2}) is applied to the qubit and 
the state of the qubit is transformed to
\begin{eqnarray}
 \sigma_y+\frac{\sigma_z}{3}~&\;{\rightarrow}\;&~\frac{10\cos\tilde{\theta}}{9}\sigma_z,
\end{eqnarray}
where $\tilde{\theta}=\arctan{(1/3})$ is chosen.
Here we consider the state of the qubit to be in its ground state, i.e.,
$\sigma_z=-1$.

From Eq.~(\ref{int_sequence}), the resulting evolution operator $\exp({Z})$
can be obtained, where $Z$ is
\begin{eqnarray}
 Z&=&\sum^{7}_{j=1}\Big[Z^{(2)}_j-\tilde{Z}^{(2)}_j\Big],\\
\label{appZ}
&=&i\hbar{\tilde{\chi}}a^\dag_1{a}_1a^\dag_2a_2a^\dag_3a_3,
\end{eqnarray}
and $\tilde{\chi}\approx{12.65}\chi_1\chi_2\chi_3\tau^3$.
Therefore, the nonlinear interactions in Eq.~(\ref{Hamiltonian}) are produced, for $k=1$.

We can ensure the convergence \cite{Suzuki} of the Zassenhaus formula if the
conditions are satisfied as
\begin{eqnarray}
 \parallel{X}_j\parallel+\parallel{Y}_j\parallel&\leq&{\ln}2-\frac{1}{2}\approx0.193,
\end{eqnarray}
We consider a subspace where the maximum excitation numbers of 
the harmonic oscillators 1 and 2 are bounded by $n_{\rm ex1}$ and
$n_{\rm ex2}$.  
The harmonic oscillator 3 is always in a coherent
state with magnitude $|\alpha|$.  Since the interactions described
in Eq.~(\ref{Ham_qubho}) do not give rise to further excitations of the harmonic oscillators,
the dynamics of the system always remains in this subspace.
Therefore, we can obtain the condition for the convergence:
\begin{eqnarray}
 2(\chi_1n_{\rm ex1}+\chi_2n_{\rm ex2}+\chi_3|\alpha|^2)\tau&\leq&0.193
\end{eqnarray}
To ensure the validity of the Zassenhaus formula,
the interaction time $\tau$ must be sufficiently short when the excitation numbers are large.
Indeed, the term, which is proportional to $\omega_z\tau$ in Eq.~(\ref{appZ}), can be cancelled by using an appropriate sequence of
rotations. Thus, a longer period of time $\tau$ can be used.

\end{document}